\documentclass[options]{JHEP3} % 10pt is ignored!
 
\usepackage{epsfig,multicol}

\newcommand\fverb{\setbox\pippobox=\hbox\bgroup\verb}
\newcommand\fverbdo{\egroup\medskip\noindent%
            \fbox{\unhbox\pippobox}\ }
\newcommand\fverbit{\egroup\item[\fbox{\unhbox\pippobox}]}
\newbox\pippobox                                                                  %
\def\){\right)}
\def\({\left( }
\def\]{\right] }
\def\[{\left[ }

\newcommand{\be}{\begin{equation}}
\newcommand{\ee}{\end{equation}}
\newcommand{\ba}{\begin{eqnarray}}
\newcommand{\ea}{\end{eqnarray}}
\newcommand{\no}{\nonumber \\}

\def\C{{\mathbb{C}}}
\def\Z{{\mathbb{Z}}}

\title{Localized tachyon condensation and G-parity conservation }

\author{Sunggeun Lee and Sang-Jin Sin \\
Department of Physics, Hanyang University, 133-791, Seoul, Korea\\
E-mail: \email{sjsin@hanyang.ac.kr }}
\preprint{\hepth{0312175}}
\abstract{We study the condensation of localized tachyon in
non-supersymmetric orbifold ${\C^2/\Z_n}$. 
We first show that the G-parities of chiral primaries are preserved under the condensation of localized tachyon(CLT) given by the chiral primaries. 
Using this, we finalize the proof of the conjecture that 
the lowest-tachyon-mass-squared increases under CLT 
at the level of type II string with full consideration of GSO projection.
We also show the equivalence between the $G$-parity given by $G=\left[jk_1/n\right]+
\left[jk_2/n\right]$ coming from partition function
and that given by $G=\{jk_1/n\}k_2 -\{jk_2/n\}k_1$ coming from 
the  monomial construction for the chiral primaires in the dual mirror picture.}

\begin{document} 
\section{Introduction}
After discovering interesting phenomena on the tachyon condensation in open string theory \cite{sen}, it has been 
tantalizing question to ask the same in closed string cases. 
The simplest closed string tachyon is the case where the closed string tachyon is 
localized at the singular point of the background geometry. In this direction, Adams, Polchinski and Silverstein \cite{aps} considered localized tachyon for non-compact orbifold $\C^r/\Z_n$ and argued that starting from a non-supersymmetric orbifolds,
there will be a cascade of tachyon condensation until space-time SUSY is restored. In a subsequent paper, Vafa \cite{vafa} reformulated the problem using the mirror picture of gauged linear sigma model, which turns out to be an orbifolded Landau-Ginzburg theory, and confirmed the result of APS.  

Since the tachyon condensation process can be considered as a renormalization goup (RG) flow\cite{harvey}, it would be interesting to ask whether there is a quantity like a c-function.  
In non-compact orbifolds, the  c-theorem\cite{zam} does not work\cite{aps,hkmm}. 
Therefore  the authors of \cite{hkmm}  tried to establish a closed string analogue of 
the g-theorem of boundary conformal field theory.  It turns out that,  if valid,  it  would give an explicit counter-example to the result of APS. On the other hand, in a related paper \cite{dv}, Dabholkar and Vafa suggested that the minimal R-charge in the Ramond sector is the height of tachyon potential at the unstable critical point.  

In \cite{namsin}, it is argued that 
the $g_{cl}$ of \cite{hkmm} does not respect the stability of supersymmetric theory and suggested a modified quantity to replace it. In a subsequent paper \cite{sin}, one of the present authors suggested that the minimal tachyon mass squared should  increase under the localized tachyon condensation. 
It turns out that this quantity is  nothing but the GSO projected version of (negative of) minimal R-charge of Dabholkar-Vafa mentioned above. 
Later, 
the statement has been studied  in a series of the papers\cite{gtheorem,fate,ring}, 
and it is proved that the R-charge decreases at the level of conformal field theories before GSO projection.
For type II theory, the proof was incomplete mainly due to the 
incomplete understanding of behavior of G-parity of GSO projection under the tachyon condensation. 
In  related papers \cite{fate} the picture of Vafa was extended  by working out the generators  of  daughter theories (the result of decay of the mother theory). The chiral rings and GSO-projection of orbifold theory was examined in more detail in \cite{ring}. 

The goal of this paper is to complete the proof of 
statement with full consideration of GSO projection,
namely to prove the following statement for type II theory of $\C^2/\Z_n$ orbifold:
\begin{itemize}
 \item Let $ m:= {\rm max}\left|\alpha'M^2\right| .$ Then, 
    $ m(UV) \geq m(IR)$, under  condensation of localized tachyon.  
\end{itemize}    

\smallskip
\FIGURE{\epsfig{file=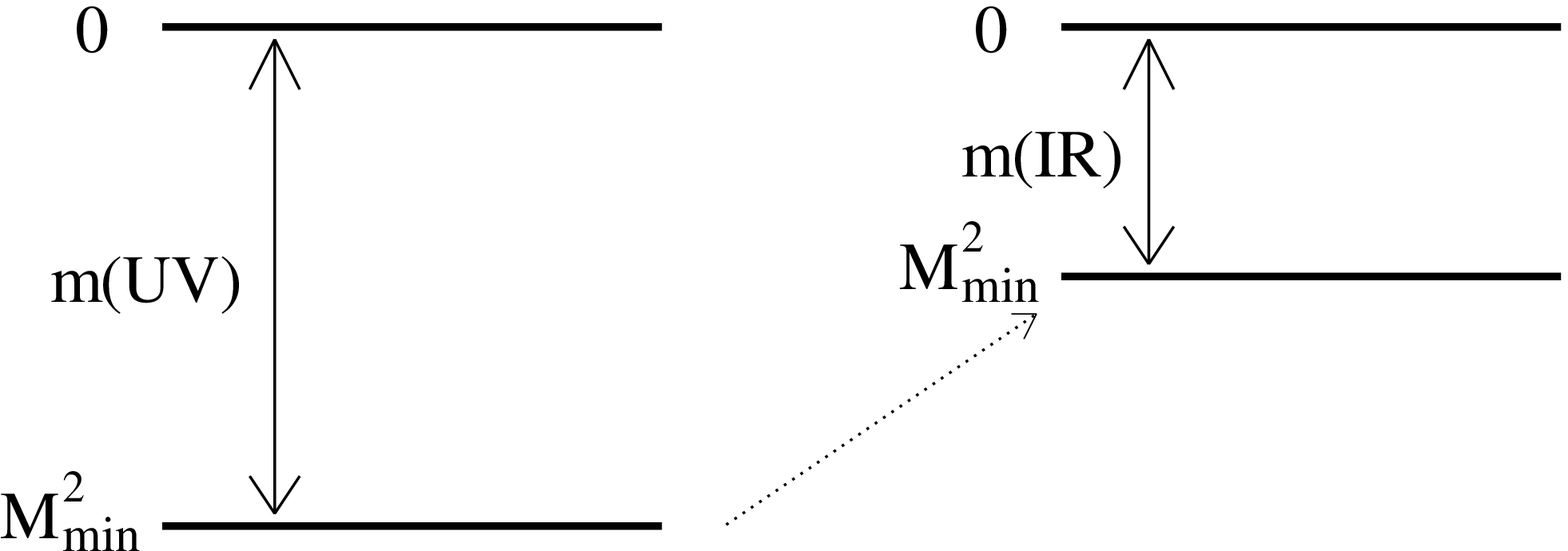,width=8cm} 
        \caption[fig1]{ $M^2_{min}$ increases and $m$  decreases under the localized tachyon condensation.}%
    \label{fig1}}

Notice that $M^2_{min}$ is negative and increases under the tachyon condensation while $m$ is positive and decreases as stated above. Figure 1 is the schematic diagram to 
clarify the content of this statement. We call this as a m-theorem to prevent possible confusion with c-theorem or g-theorem.  

The rest of the paper goes as follows. In section 2, 
we consider  the consistency  of various GSO-projections introduced by arbitrarily different authors using different logics. In section 3, we show that,  $n(k_1,k_2)$, the $\Z_n$ orbifold with  generator $(k_1,k_2)$  equivalent to $n(1,k)$ for some $k$ which we will fix in detail. We call the latter as  "canonical representation".
In section 4, we prove that the $G$ parity is conserved under the condensation  of the localized tachyon given by the chiral primaries. This is the most important step in proving the m-theorem. 
In section 5, we finish the proof of the m-theorem. In section 6,  we give a 
discussion on the implication of the theorem and  conclude. 

\section{Equivalence of various GSO-projections}

In this section we first want to understand whether orbifold
GSO chiral projections
recently introduced in \cite{vafa,hkmm,sin}
are mutually consistent.
Let $k_1,k_2$ be the generator of the orbifold action of $\Z_n$, that is,
\be
x^{(1)}(z) \to e^{2\pi ik_1/n }x^{(1)}, \;\; x^{(2)}(z) \to e^{2\pi ik_2/n }x^{(2)} ,\ee
where $x^{(1)},x^{(2)}$ are complex co-ordinate of $\C^2$.
We  represent $\C^2/\Z_n$  with generator  $(k_1,k_2)$ by $n(k_1,k_2)$.
In HKMM \cite{hkmm}, $k_1=1$ cases were discussed. Here we discuss  their result in the extended form, that is,  with general $(k_1,k_2)$. The chiral 
primary operators were constructed from
bosonized world sheet fermions  $\psi_i =e^{iH_i}$ as
\be
X_j=X^{(1)}_{n\{ {jk_1 \over n}\} }X^{(2)}_{n\{ {jk_2 \over n} \} } ,
\ee
where
\be
X_j =\sigma_{j\over n}e^{i {j\over n}(H-\bar{H})},
\ee
with $\sigma_{j\over n}$ being a twist operator. The $Z_2$ action defining the GSO projection is given by
\be
H_1\to H_1 + k_2 \pi, \quad H_2 \to H_2 -k_1 \pi.
\ee
In untwisted sector it acts as $(-1)^{F_L}$ and
restricts both $k_1$ and $k_2$ to be an odd. In twisted sector $X_j$ has
phase
\be
X_j\to e^{i\pi (k_2 \{ {jk_1\over n}\}-k_1 \{ {jk_2 / n} \} )}X_j
:=(-1)^s X_j,
\ee 
where $\{x\}=x-[x]$ with 
$[x]$ being the  greatest integer that does not exceed $x$.
 Note that $s$ is an integer in general and especially when $k_1=1$ and $k_2=k$ 
\be
s=\[{jk \over n}\].
\ee

In \cite{vafa}, Vafa reformulated the orbifold problem as Landau-Ginzburg theory by imbedding the orbifold geometry in the
gauged linear sigma model\cite{glsm} and subsequently taking the  mirror dual. The superpotential coming from the vortex contribution can be written as
\be
W= u^n_1 + u^n_2 + e^{{t/ n}} u^{p_1}_1 u^{p_2}_2
\ee
$(-1)^{F_L}$ should be defined by requiring $W \to -W $.
This can be achieved by defining the $Z_n$ action on $u_i$ by
\footnote{We have to make modifications of discussion of \cite{vafa} on GSO projection.}
\be
u_1 \to e^{i \pi {k_2 /n} }u_1, \quad u_2 \to e^{-i\pi {k_1 / n} }u_2.
\label{uzn}
\ee
As a result,
\be u_1^{p_1}u_2^{p_2}\to
(-1)^{p_1 {k_2/ n}-p_2 {k_1 / n}}u_1^{p_1}u_2^{p_2}:=(-1)^{s'}
u_1^{p_1}u_2^{p_2}.
\ee
So by identifying $p_i = n\{ jk_i/n\}, \;\; i=1,2,$
we get 
\be s= p \times k /n=s'
,\ee
and consequently, two GSO actions are completely consistent.
Notice that $s$ is always an integer.
We also see that  $u_1^{p_1}u_2^{p_2}$ in the mirror LG theory corresponds to $X_j$ of the 
operator construction.
%\footnote{If we define GSO by exchanging $k_1$ and $k_2$ in
%eq.(\ref{uzn}), we still have $u_i^n \to -u_i^n$ but ${ p\cdot k/n}$,
%the phase of $u_1^{p_1}u_2^{p_2}$, will not be an integer in general,
%hence GSO projection is not well defined. }

We now want to see whether 
the GSO projection coming from the partition function \cite{sin,ring} is also consistent with above two. 
The result of ref. 
\cite{sin} shows that $u_1^{p_1}u_2^{p_2}$ is projected out if
\be
G=[jk_1/n]+[jk_2/n]
\ee
is even (odd) for $cc$ ring ($ac$ ring).
For our purpose, it is enough to show that
\be
G \equiv s \; {\rm mod}\; 2. \label{mod2}
\ee
We first notice that for special case $k_1=1, k_2=k$,
\be
s=[jk/n]=G.
\ee
In the next section we will prove that 
 all $n(k_1,k_2)$ have equivalent representation $n(1,k)$ for some $k$. 
 So the above result  is  enough if we consider only a given theory. 
However, we will need to consider the case where $k_1\neq 1$ when we consider the `decay'  of  $n(1,k)$ by explicitly specifying  the daughter theories. So let's consider the general cases. 
For type II, we need to have $k_1 + k_2 =even$ \cite{ring}.
If both $k_1,k_2$ are odd integers,
the equivalence can be readily seen by considering
$s+G$ with the  help of the following identity.
\be
s=k_2 \{ jk_1 /n \} -k_1 \{ jk_1/n \} =-k_2[jk_1/n]+k_1[jk_2/n],\ee
so that $s+G$ is even, which is enough for our goal. 
If both $k_1,k_2$ are even, then $s$ is even and $G$ is not necessarily equivalent to $s$. In table \ref{n5}, we give an  explicit example for this case.
However, if we further restrict ourselves to the case 
where  $n,k_1,k_2$ are mutually co-prime, 
we can restrict ourselves to the case where $k_1, k_2$ are  odd. 
Later in section 4, we will only need to consider the case where $k_1=1$ or both $k_1,k_2$ are odd.

\begin{table}
\centering
\begin{tabular}{|c|c|c|c|c||c|c|c|c|c||c|c|c|c|c|}
 \hline
 $j$ & $cc$ & $G$ & $s$ & $ca $  &  $j$ & $cc$ & $G$ & $s$ & $ca $  &$j$ & $cc$ & $G$ & $s$ & $ca $   \\
 \hline
 $1$ & $(2,4)$ & $0$ & $0$ & $(2,1)$  &$1$ & $(1,2)$ & $0$ & $0$ & $ (1,3)$   &$1$ &$(1,2)$ & $-1$ & $-1$  & $ (1,3)$  \\
 $2$ & $(4,3)$ & $1$ & $2$ & $(4,2)$  &$2$ & $(2,4)$ & $0$ & $0$ & $ (2,1)$   &$2$ &$(2,4)$ & $-2$ & $-2$ & $ (2,1)$  \\
 $3$ & $(1,2)$ & $3$ & $0$ & $(1,3)$  &$3$ & $(3,1)$ & $1$ & $1$ & $ (3,4)$   &$3$ &$(3,1)$ & $-2$ & $-2$ & $ (3,4)$  \\
 $4$ & $(3,1)$ & $4$ & $2$ & $(3,4)$  &$4$ & $(4,3)$ & $1$ & $1$ & $ (4,2)$   &$4$ &$(4,3)$ & $-3$ & $-3$ & $ (4,2)$    \\
 \hline
\end{tabular}
\caption{\scriptsize Comparison of  $G$ and $s$ in  $\bf 5(2,4)$(left) $\bf 5(1,2)$(middle) and $\bf 5(1,-3)$(right).
$\bf 5(2,4)$(left) provides an example where $s \neq G$ mod 2. However, $s \equiv G$ mod 2  for all elements 
in  $\bf 5(1,2)$ and $\bf 5(1,-3)$ . \label{n5}}
\end{table}

\section{Equivalence of $n(k_1,k_2)$ and $n(1,k)$}

 Now, we want to show that $n(k_1,k_2)$ is  equivalent with  $n(1,k)$ for some $k$. 
We can choose a convention where  $k_1>0$, since $n(k_1,k_2) = n(-k_1,-k_2)$ 
even after GSO projection. \footnote{One can see this from $[-jk_i/n]=-[jk_i]-1$ 
regardless of the sign of $k_i$.}
First, notice that $(k_1,k_2)$ and $(1,k)$ should generate the same spectra if  
$k_2/k_1=k$ and $n,k_1$ and $n,k_2$ are
relatively co-prime, since the spectrum is nothing but the
modulo-$n$-rearrangement of $j(k_1,k_2)$ for $j=1, \cdots, n-1$.
In fact, any of the element of the spectrum, that is, any of
$j(k_1,k_2)$ modulo $n$ can be the generator of the same spectrum
set. Therefore, without GSO projection, the equivalence of
the two is quite obvious.
For type 0 case, the GSO projection does not eliminate any variety of chiral 
primaries in the following sense:
if an operator with a certain charge is projected out in $cc$ ring, 
there is a surviving operator in $aa$ ring 
with the same charge. 
This can be seen from the fact that $j$-th element of 
$cc$-ring and $(n-j)$-th element of $aa$ ring have the same charge 
 but different G-parity if $k_1+k_2=odd$ \cite{ring}:
 \be
n(1-\{(n-j)k_i/n\})=n\{jk_i/n\}, \;\;\;  G(n-j) \equiv G(j)+k_1+k_2 \;{\rm mod}\;2.\ee 
 Similar relations hold between $ca$ and $ac$ rings. 
Therefore if two type 0 theories have the same spectrum  before GSO projection, so do they after GSO.
 
Hence from now on,  we concentrate on the type II case, where $k_1+k_2=even$. 
We first have to specify  $k$ more precisely. Let $k_1^{-1}$ be the
multiplicative inverse of $k_1$ in $\Z_n$ so that there is a
unique integer $a$ depending on $k_1$ such that 
\be k_1^{-1}k_1=na+1, \;\; \hbox{for any given } k_1. \label{a}\ee 
Then, $k$ is equal to $k_1^{-1}k_2$ modulo $n$.
So there exists  an integer $l$ such that 
\be k= k_1^{-1}k_2+ln , \;\; {\rm and}\;\; -n<k<n \label{k}.\ee 
In fact, there are two such $l$'s, since the length of range is $2n$. They are consecutive. 
In order for $n(1,k)$ to be a type II string theory, we require that 
\be
k=k_1^{-1}k_2+ln  =odd.\ee  
For  odd $n$, this fixes $l$ uniquely, since $k \neq k \pm n$ mod 2. 
However, for even $n$, the ambiguity  will be removed 
only after we take account the G-parity of $k^{-1}$ and $l$ more carefully.  
Before we proceed, we give some examples to give
some feeling on how things work.

\begin{itemize}
    \item $8(3,-5)$: It is the same with $8(1,1)$ before GSO projection, 
    since $-5\cdot 3^{-1}\equiv 1$ in $\Z_8$.  However, these are NOT equivalent
     after GSO as one can see from the table \ref{n8}. Then, it may look like 
       a counter example. However,  for even $n$, both $n(k_1,k_2)$ and   $n(k_1,k_2\pm n)$ 
represent the same   type of theory regarding  to whether they are type 
0 or type II \cite{ring}. Here $\pm$ is  chosen such that $-n<k_2 \pm n<n 
$ is satisfied.
Therefore we should also consider $8(1,-7)$ instead of $8(1,1)$.  
Remarkably, $8(3,-5)$ and $8(1,-7)$ have the same GSO projected spectrum as 
one can see from table \ref{n8}.  

\item  $7(3,5)$: It is equivalent to  $7(1,-3)$  and also to $7(1,4)$
before GSO projection. But $7(3,5)$ and $7(1,-3)$ is a type II
while $7(1,4)$ is type 0. Therefore in this case there is a unique representation in the same type.  One can explicitly check that $7(3,5)$ has  identical spectrum  with
$7(1,-3)$ after GSO from table \ref{n7}.
 
\end{itemize}

\begin{table}
\centering
\begin{tabular}{|c|c|c|c|c||c|c|c|c|c||c|c|c|c|c|c|}
 \hline
$j$ & $cc$  & $G$ & $s$ &$ca$   &$j$ &$cc$  & $G$& $s$  & $ca$ &$j$ &$cc$ & $G$ & $s$ & $ca$   \\
 \hline
$1$ & $(5,5)$ & $-1$& $5$& $(5,3)$   &$1$ &$(1,1)$ & $0$&$0$ & $(1,7)$ &  $1$ &$(1,1)$ & $-1$&  $-1$& $(1,7)$  \\
$2$ & $(2,2)$ & $0$ & $2$& $(2,6)$   &$2$ &$(2,2)$ & $0$&$0$ & $(2,6)$ &  $2$ &$(2,2)$ & $-2$&  $-2$& $(2,6)$  \\
$3$ & $(7,7)$ & $-1$& $7$& $(7,1)$   &$3$ &$(3,3)$ & $0$&$0$ & $(3,5)$ &  $3$ &$(3,3)$ & $-3$&  $-3$& $(3,5)$  \\
$4$ & $(4,4)$ & $0$ & $4$& $(4,4)$   &$4$ &$(4,4)$ & $0$&$0$ & $(4,4)$ &  $4$ &$(4,4)$ & $-4$&  $-4$& $(4,4)$ \\
$5$ & $(1,1)$ & $1$ & $1$& $(1,7)$   &$5$ &$(5,5)$ & $0$&$0$ & $(5,3)$ &  $5$ &$(5,5)$ & $-5$&  $-5$& $(5,3)$ \\
$6$ & $(6,6)$ & $0$ & $6$& $(6,2)$   &$6$ &$(6,6)$ & $0$&$0$ & $(6,2)$ &  $6$ &$(6,6)$ & $-6$&  $-6$& $(6,2)$  \\
$7$ & $(3,3)$ & $1$ & $3$& $(3,5)$   &$7$ &$(7,7)$ & $0$&$0$ & $(7,1)$ &  $7$ &$(7,7)$ & $-7$&  $-7$& $(7,1)$   \\
 \hline
\end{tabular} 
\caption{\scriptsize $cc$-,$ca$-ring of 
$\bf 8(-3,5)$ (left), $\bf 8(1,1)$ (middle), $\bf 8(1,-7)$  (right). $G,s$ for each element are given for comparison. Both
 $8(1,-7)$ and $8(1,1)$ are equivalent to $8(-3,5)$ before GSO.
 But only $8(1,-7)$ is so after GSO. This is a general phenomena: For even $n$ type II, $k$ is not determined uniquely from $(k_1,k_2)$ before GSO.  This ambiguity or freedom will be essential to find correct $k$ with GSO projection considered.}
\label{n8}
\end{table}

\begin{table}
\centering
\begin{tabular}{|c|c|c|c||c|c|c|c|}
 \hline
$j_1$ & $cc$-elements & $G$ &$ca$-elements &$j_1$ & $cc$  & $G$ &$ca$    \\
 \hline
$1$ & $(3,5)$ & $0$& $(3,2)$ &$1$ & $(1,4)$ & $-1$& $(1,3)$  \\
$2$ & $(6,3)$ & $1$ & $(6,4)$& $2$ & $(2,1)$ & $-1$ & $(2,6)$ \\
$3$ & $(2,1)$ & $3$& $(2,6)$ & $3$ & $(3,5)$ & $-2$& $(3,2)$ \\
$4$ & $(5,6)$ & $3$ & $(5,1)$ & $4$ & $(4,2)$ & $-2$ & $(4,5)$ \\
$5$ & $(1,4)$ & $5$ & $(1,3)$ & $5$ & $(5,6)$ & $-3$ & $(5,1)$ \\
$6$ & $(4,2)$ & $6$ & $(4,5)$ & $6$ & $(6,3)$ & $-3$ & $(6,3)$ \\
 \hline 
\end{tabular}
\caption{\scriptsize $cc$-, $ca$-rings for $\bf 7(3,5)$ (left) and 
$\bf 7(1,-3)$ (right). Notice that 7(1,4) is a type 0 theory and it is not tabulated here. 
For odd $n$ type II, $k$ is chosen uniquely from $(k_1,k_2)$.\label{n7}}
\end{table}

Now let us come back to the general argument.
Consider $G$-parity for $n(k_1,k_2)$ and $n(1,k)$ with $k$ given in eq.(\ref{k}). 
We call them as $A$ and $B$ orbifold
theory respectively.  The $G$-parity for $j$-th element of $A$ is
$G_A(j)=[jk_1/n]+[j k_2/n]$ and  the $G$-parity  for $j$-th one of $B$ is $G_B(j)
=[jk/n]$.  We remind that the $cc$- and
$ca$-rings of $A$ and $B$ are the same (as sets) before  GSO projection.
Let the $j$-th element of A theory appears as the $j'$-th element for B so that  \be j'=n\{jk_1/n\}. \ee
 For our purpose, it is enough to show  that 
\be
G_A(j)\equiv G_B(j')\; {\rm mod} \; 2.\ee 
Since $k$ is an odd integer,  
 \be
G_B= [(jk_1/n- [jk_1/n]) k] \equiv [jk_1/n]+jk_1(k_1^{-1}k_2+ln)/n \;\;{\rm mod}\;2. \label{Gab}\ee
Using eq.(\ref{a}), one can easily show that  
\be
G_B \equiv G_A+jk_1l+jk_2a \;\;{\rm mod}\;2.
\ee
If both $k_1,k_2$ are even and $n$ is odd, then  our job is done.
If both are odd, then $G_B \equiv G_A+j(l+a)$ modulo 2. 
For even $n$, one of the consecutive $l$'s (see below eq.(\ref{k})) can 
be chosen such that $l+a$ is even and this 
condition removes the ambiguity in the choice of $l$ (hence in  $k$) as 
mentioned before. 

For odd $n$, $l$ is fixed as follows:
If $k_1^{-1}k_2$ is already odd, then $l$ should be even not to change the type 0/type II. 
If the former is even, then $l$ should be odd. For $k_1^{-1}k_2$ odd case, using eq. (\ref{a}), 
\be
(na+1) k_2=k_1\cdot odd \equiv k_1\; {\rm mod}\; 2.
\ee
If  $k_1, k_2$ are both odd, 
$na+1$ should be odd. Since $n$ is odd, $a$ must be even. Therefore if $k_1^{-1}k_2$ is odd and $k_1,k_2$ have the same number of factor 2, $l+a$ is even as desired.
Similarly,  for $k_1^{-1}k_2$ even case, 
\be
(na+1) k_2=k_1\cdot even \equiv k_1\; {\rm mod}\; 2.
\ee 
If $k_1,k_2$ are   both odd, then $na+1$ must be even, hence $a$ must be even. 
Therefore in this case also $l+a$ is even as desired.
Hence we proved the following \begin{itemize}
    \item {\it {\bf Lemma:} 
For any orbifold $n(k_1,k_2)$, we can represent it by $n(1,k)$ with GSO projection properly considered.}
\end{itemize}
 
\section{Conservation of $G$-parity under LTC}
The final most important step in proving the m-theorem is to show that  $G$ parity is conserved under the localized tachyon condensation.  Namely, 
if a particular charge is projected out in the mother theory, its daughter image under $T_p$ is also projected out in a daughter theory. 

In the previous work \cite{fate}, we showed that 
the decay of $n(k_1,k_2)$ under the condensation of localized tachyon with weight $p=(p_1,p_2)$ is
\be
n(k_1,k_2) \to p_1 (k_1,s)\oplus p_2(-s,k_2) \label{decay}
\ee
where $p_1= \{ jk_1 /n\} $, $p_2 =\{ jk_2/n\} $ and
$s= p \times k /n= k_2 \{ jk_1 /n \} -k_1 \{ jk_2 /n\} $.
Its $G$-parity is given by
\be
G_p=[ {jk_1 \over n } ]+ [ {jk_2 \over n} ] \quad {\rm mod}\; 2
\ee
Here $[x]$ means the integer part of $x$ while $\{ x \}$ means the fractional
part of $x$. In order for $(p_1,p_2)$ to survive, $G_p$ should be
an odd integer.  
On the other hand, by the result of last section, we only need to consider 
the decay of a canonical representation:
\be
n(1,k) \to p_1(1,s_p)\oplus p_2(-s_p,k), \label{process}
\ee
with  $p_i=n\{jk_i/n\}, s_p=p\times k/n=(p_2-kp_1)/n$.
In order to fix the ambiguity in the daughter-theory-generators, we use the fact that  
for type II theory, the bulk tachyon is projected out and it can not and should not be regenerated 
by the localized tachyon condensation process. That is, type II can decay only to type II.
Then $s$ and $k$ should be both odd. This conditions are already satisfied:
the orginal theory is type II hence $k$ is odd and $p$ as a surviving 
element of $cc$-ring must have odd $G_p(=s_p)$.
If any of $p_1$, $p_2$ is odd, that can not be added or subtracted to the generator,
since it convert the type II to type 0. 
If any of them are even, it can be added to the generator and we have ambiguity in the 
determination of the daughter theory. 
However, as we will see shortly, this can not be so and we will see that the eq. (\ref{process}) 
describe the tachyon condensation properly even after the GSO projection.
For a moment, we assume that eq. (\ref{process}) is true.

We need the map which gives the G-parity value $[lk/n]$ when the data $(q_1,q_2)=(l,n\{lk/n\})$ of a
charge are given.  This is given by the observation:
 \be
 G_q=[lk/n]=(kq_1-q_2)/n = q\times k/n. \label{qtoG}
 \ee
Under the  condensation  of $(p_1,p_2)$, $q$ is mapped to $q'$ by
the tachyon map
$T_p$ \cite{gtheorem}. If $q'$ belongs to up-theory of daughter theory,
\be
q'=T^+_p q=\pmatrix{ 1 & 0 \cr
          -p_2/n & p_1/n }\pmatrix{ q_1 \cr q_2 }
=\pmatrix{ q_1 \cr p\times q/n}
=\pmatrix{q_1' \cr q_2'}.
\ee
Then,  
\be
G_{q'}={ q_2'-q_1's_p \over p_1}
={1\over {np_1}} (p\times q -q_1(p\times k))
={1\over n}(q\times k)
=G_q,
\ee
which proves the conservation of G-parity for the up-theory spectrum. 
Similarly we can prove the same statement for the down-theory using  
the equivalence of G-parity with $s$ proved before. Notice that both $s$ and $k$ are odd,
so that the result applies here. Since
\be
q''=T^-_p q =
( - p\times q/n, q_2):=(q_1'', q_2''),\ee
we have 
\be
G_{q''}={ q_2''s+q_1''k  \over p_2}
={1\over {np_2}} (-(p\times q)k +q_2(p\times k))
={1\over n}(q\times k)
=G_q,
\ee
finishing the proof of G-parity conservation.
\footnote{We can also use the inverse representation $
n(k',1) \to p_1(k',s'_p)\oplus p_2(-s'_p,1),$
to prove the G-parity conservation for the down theory spectrum. Here
$k':= k^{-1}$ in the finite field $\Z_n$, and $s'_p= p\times k'/n=
(p_1-k'p_2)/n$.
Using  $G=[lk'/n]=(q_2k'-q_1)/n$, we have 
$G_{q'}={ q_2'(-s'_p)-q_1 \over p_2}
={1\over {np_2}} (-q_2(p\times k')+p\times q )
={1\over n}(q_2k'-q_1)
=G_q.$ 
}

An important remark is in order.
The decay rule eq. (\ref{decay}) or  eq.(\ref{process}) is given at the conformal field theory 
level. After GSO, since $n(k_1,k_2)$ and $n(k_1,k_2\pm n)$ are different theories, 
it seems that the decay rule has the same ambiguity: 
that of adding $\pm p_1$, $\pm p_2$ to $s$ and $k_2$ respectively in the right hand side of 
eq. (\ref{decay}): why the following process is not allowed? 
\be
n(k_1,k_2) \to p_1 (k_1,s\pm p_1)\oplus p_2(-s,k_2\pm p_2) \label{modified}. 
\ee  
If  $p_1(k_1,s_p)$ is replaced by $p_1(k_1,s_p \pm p_1)$, then a 
computation shows that 
\be
G_{q'}=G_q \mp q_1.
\ee 
This means that an operator that is projected out in the mother theory can 
be  resurrected in daughter theory under the localized tachyon condensation(LTC).
This is not physical, because what changes under the LTC is not the operator $u_1^{q_1}u_2^{q_2}$
but the Lagrangian: 
\be W=  u_1^{n}+ u_2^{n}+e^{t/n} u_1^{p_1}u_2^{p_2} .\label{orLGeq}
\ee
Namely, each operator remains the same and  only the coefficient $e^{t/n}$ of 
condensing tachyon operator changes  so that the  
measuring method of the weight of operators $u_1^{q_1}u_2^{q_2}$ 
changes. 
Once an operator is projected out at the moment of $t\to -\infty$, 
it is not possible  to resurrect it  by a smooth deformation of taking $t\to \infty$.
The same argument holds for the $ p_2(-s_p,k\pm p_2)$.  Therefore the original rule eq.(\ref{decay}) or 
eq.(\ref{process}) remains its form and 
the modified rule eq.(\ref{modified}) is not allowed.  

\section{Proof of m-theorem}
According to the result of \cite{gtheorem}, the spectrum of 
daughter theories can be regarded as images of certain linear mapping $T$ for $cc$ ring  and $F$ for $ca$ or $ac$ rings
given by 
\ba
T^+_p(q)=(q_1,p\times q/n), & T^-_p(q)=(-p\times q/n,q_2), \no
F^+_p({\bar q})=(q_1,p_1-p\times q/n), & F^-_p({\tilde q})=(p_2+p\times q/n,q_2).
\ea
Here, superscript $+$ is for up-theory, $-$ is for down-theory and ${\bar q}=(q_1,n-q_2) \in ca$ ring, ${\tilde q}=(n-q_1,q_2) \in ca$ and ${p'}=(p_1,-p_2)$.
$T$ is a mapping for tachyon condensation which exists due to the worldsheet ${\cal N}=2$ SUSY, while $F$ is 
not a mapping describing tachyon condensation and its existence is due to the conformal symmetry of the final theory. For more detail, see \cite{gtheorem}. 
The most important property of $T$ and $F$ is their monotonicity,
which enables the proof of m-theorem possible.

The conservation of $G$ parity means that 
the same logic  can be applied to  the spectrum of type II even after GSO projection.  The worst thing that can happen is the case where the  operator with minimum R-charge  of the Mother theory is deleted in the mother theory but resurrected in the daughter theory. The G-parity conservation guarantees that such phenomena can never happen. See figure \ref{fig2}.
\smallskip
\FIGURE{\epsfig{file=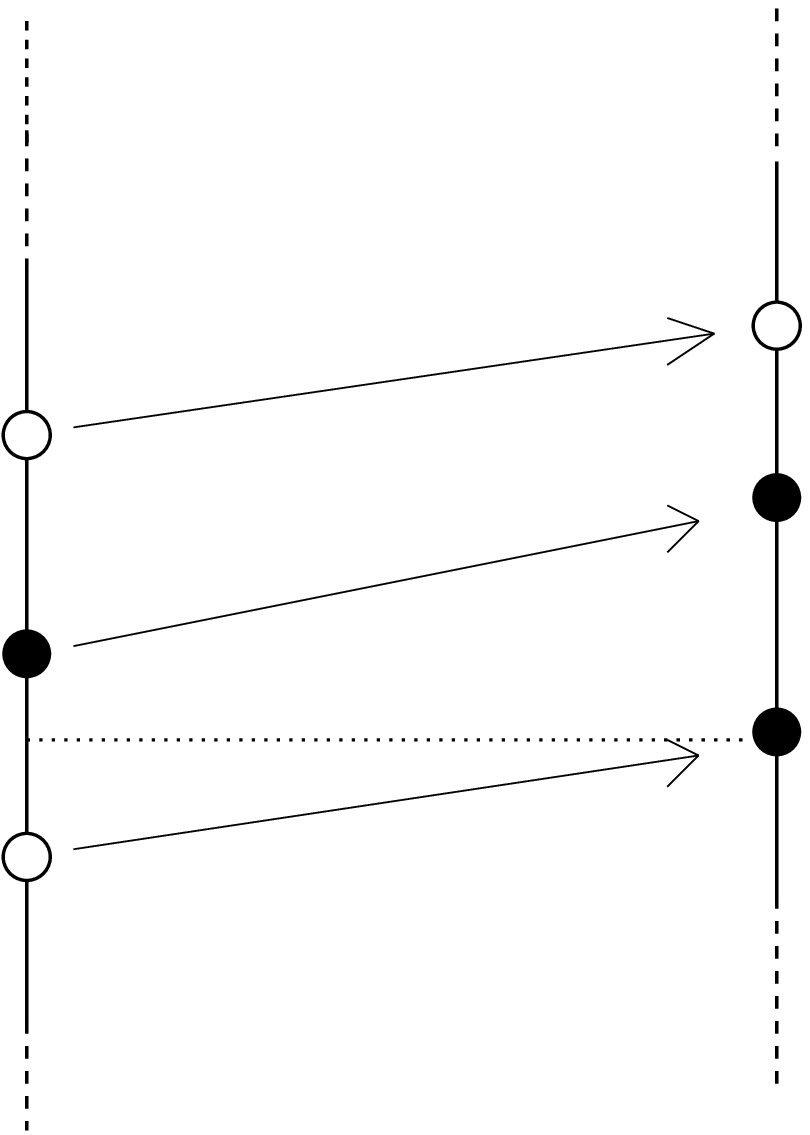,width=6cm} 
        \caption[fig1]{A candidate phenomena that can destroy the m-theorem in the absence of the G-parity conservation. Filled dots represent operators surviving under the GSO projection and empty dots are those projected out.}%
    \label{fig2}}

For completeness, we describe  some of the detail here. Let $p$ be the condensing tachyon and  
$q'_{min}$ be the element with minimal R-charge of the daughter theory,  i.e. the smallest one among  all charges in up and down theories.
If $q'_{min}$ belongs to  $cc$ ring of up-theory, then it is an image of an element $q$ in the mother theory under $T^+_p$. 
$q$ must belong to the $cc$-ring of the mother theory due to the G-parity 
conservation. If $q_{min}$ is the minimal element of 
the mother theory, $R[q_{min}]$, the R-charge of $q_{min}$, is smaller than  $R[q]$ by definition. Then our desired statement, 
\be
R[q_{min}]\le R[q'_{min}]
\ee
comes from the monotonicity of $T^+$, namely,
\be
 R[q]\leq R[T_p^+(q)]=R[q'_{min}].\ee
If $q'_{min}$ belongs to  $cc$ ring of down-theory,
then  $T^-_p$ replaces $T_p^+$ for above argument.
If $q'_{min} \in ca$-ring of daughter theories, then 
the entire arguments can be repeated by replacing the map  $T^\pm$ by $F^\pm$ to finish the proof.

\section{Discussion}
Our proof was actually motivated from the numerical work we performed, 
which showed that the theorem holds for all $n(k_1,k_2)$ with $n\leq 100$ and 
$-n<k_1<k_2<n $, which provided significant evidence to believe that the m-theorem is true. 

The m-theorem we just proved implies that the one loop cosmological constant of the non-supersymmetric orbifold is a 
monotonically decreasing quantity under the localized tachyon condensation. This is because, 
the former is defined as the integral of one loop partition function over the fundamental domain with some cutoff to control the divergence. The main divergence comes from the low temperature limit of the partition function   which is dominated by the contribution of the lowest tachyon mass.
Therefore the present work proves the statement of the original conjecture of \cite{sin} in its full strength.
The cosmological constant vanishes when the theory  reaches a supersymmetric point by the tachyon condensation.

We describe some future work. 
Our work is still comparing two end points of RG where conformal invariance makes the explicit computations available.
Therefore it is still needed to extend our work to the string off-shell level, 
which would require much more non-trivial efforts. 
Some preliminary work in this direction for the simplest case  is discussed in Dabholkar and Vafa \cite{dv}. Extension in this direction is in progress \cite{leesin2}. 
 
\bigskip

\acknowledgments

This work is supported by KOSEF Grant 1999-2-112-003-5.

\end{document}